\documentclass[aps,twocolumn,showpacs]{revtex4}
\usepackage{graphicx}
\include{amssym}

\newcommand{\ud}{\mathrm{d}}

\begin{document}

\title{The $\pi^0-\eta-\eta'$ mixing in a generalized multi-quark interaction scheme.}

\author{A. A. Osipov\footnote{Email address: osipov@nu.jinr.ru},
        B. Hiller\footnote{Email address: brigitte@teor.fis.uc.pt}
    and A. H. Blin\footnote{Email address: alex@teor.fis.uc.pt}}
\affiliation{CFisUC, Department of Physics, University of Coimbra, 3004-516 Coimbra,
Portugal}

\begin{abstract}
We investigate the isospin symmetry breaking effects within a recently
derived Nambu-Jona-Lasinio related model by fitting the measured pseudoscalar
meson masses and weak decay couplings $f_\pi$, $f_K$. Our model contains the next
to leading order terms in the $1/N_c$ expansion of the effective multi-quark Lagrangian,
including the ones that break the chiral symmetry explicitly. We show the
important phenomenological role of these interactions: (1) they lead to an
accurate fit of the low-lying pseudoscalar nonet characteristics; (2) they
account for a very good agreement of the current quark masses with the present PDG values;
%with the celebrated current algebra
%Weinberg's result $m_u/m_d=0.56$, $m_s/m_d=20.1$ for the light quark masses; 
(3) they
reduce by $40\%$ the ratio $\epsilon/\epsilon'$ of the $\pi^0 -\eta$ and $\pi^0 -\eta'$ mixing angles, as compared to the case that contemplates explicit breaking only in the leading order, 
bringing it in consonance with the quoted values in the literature. The
conventional NJL-type models fail in the joint description of these
parameters.

%Isospin symmetry breaking effects related to the  $\pi^0-\eta$ and $\pi^0-\eta'$ mixing angles ${\epsilon},{\epsilon'}$ are considered within a recently derived multi-quark interaction Lagrangian which accounts for all possible spin zero non-derivative type of vertices relevant at the scale of spontaneous breaking of chiral symmetry in 4D, including a subset of interactions which break the chiral symmetry explicitly, apart from the standard QCD mass term.  %We work in the strange non-strange basis where the interactions of the $\pi^0$ with the $\eta$ and $\eta'$ mesons can be linearized in the mixing angles. 
%We obtain a reduction of $40 \%$ in the ratio of the mixing angles $\frac{\epsilon}{\epsilon'}$ when the explicit symmetry breaking terms are present, as compared to  bringing it in consonance with the quoted values in the literature. Furthermore a very good agreement with the empirical values for the current quark masses $m_u,m_d$ is achieved together with an accurate description of the low lying pseudoscalar spectrum.  
\end{abstract}

%Keywords: QCD, Explicit symmetry breaking, Effective theory, Mesons.
\maketitle

{\bf I. Introduction}
\vspace{0.5cm}
 
The strong isospin symmetry is considered to be a very good approximation in the empirical description of a large bulk of strong interaction processes. This is related to the hierarchy in which  breaking of the chiral symmetry $SU(3)_L \times SU(3)_R$ by different current quark masses occurs, down to $SU(2)_I \times U(1)_Y$ flavor symmetry if $m_u,m_d \ll m_s$. In the case of the pseudoscalar mesons it is accurate at the order of the ratio of the light and strange current quark masses $(m_u-m_d)/m_s$ \cite{Gross:1979},\cite{Gasser:1985} and explains partly the small meson mass differences within charged isospin multiplets. A further source of isospin breaking is due to the electromagnetic interactions, which are expected to be suppressed at the scale of strong interactions. 

A detailed quantitative analysis however requires isospin breaking corrections to be taken into account in a series of low energy phenomena, such as: the description of mass splittings of mesons, and Dashen's theorem \cite{Urech:1995};  sum rules for quark condensates \cite{Gasser:1985}, \cite{Nicola:2010};  kaon decays \cite{Nehme:2004}; $\pi-\pi$ \cite{Meissner:1997},\cite{Knecht:1998} and $\pi-K$ scattering \cite{Nehme:2002},\cite{Kubis:2002} in relation  to mesonic atoms, \cite{Knecht:2002},\cite{Schweizer:2004}, $\rho$ and $\tau$ decays involving $\eta(\eta')$ mesons \cite{volkov:2012},  $a_0-f_0$ mixing \cite{Hanhart:2007} in the scalar meson sector.

Strong isospin breaking effects become particularly relevant if a certain process depends crucially on the differences of the light quark masses. If in addition the electromagnetic interactions are a subleading effect, these processes provide for ideal tools in a quantitative analysis of quark mass ratios. In the latter category are the $\eta,\eta' \rightarrow 3 \pi$ decays, the $\pi^0-\eta$ and $\pi^0-\eta'$ mixings, as well as  the $\rho-\omega$ mixing in the vector channels. 

Isospin breaking associated with the $\pi^0-\eta-\eta'$ system has long been known to play a role in the Standard Model prediction of the CP violation related ratio $(\frac{\epsilon'}{\epsilon})_{CP}$ \cite{Bijnens:1984,Donoghue:1986,Buras:1987,Cheng:1988} representing a substantial correction to the QCD penguin contributions \cite{Buras:1987}. It affects the value of the $K^0\rightarrow \pi^0\pi^0$  transition through the dominant QCD $Q_6$ penguin operator, which is one of the sources of uncertainties in the determination of $(\frac{\epsilon'}{\epsilon})_{CP}$ \cite{Buras:1993}, for a recent review see \cite{Buras:2014}. 

In chiral perturbation theory (ChPT)\cite{Weinberg:1979},\cite{Gasser:1984},\cite{Gasser:1985} the $\pi^0-\eta$ mixing angle occurs already at order $p^2$ and was first evaluated to order $p^4$ in the context of $K_{l3}$ form factors in \cite{Gasser:1984}. 

In the analysis of $\eta-\eta'$ mixing of \cite{Feldmann:1998}  the $U(1)_A$ anomaly is described by the gluon transition matrix element $<0|\frac{\alpha_s}{4\pi}$G\~G$|\eta_i>$ and the quark flavor basis has been used. As shown in \cite{Schechter:1993} this basis is favored, as one of the two mixing angles is indeed small. The decay constants follow the pattern of particle state mixing in that basis. It has been shown that this approach leads to results consistent with many observables related to $\eta-\eta'$ mixing \cite{Feldmann:1998}. 
In \cite{Feldmann1:1999},\cite{Kroll:2005} it has been extended to include the mixing to the neutral pion.

In the present work we address the $\pi^0-\eta-\eta'$ mixings resulting from a recently proposed Lagrangian \cite{Osipov1:2013},\cite{Osipov2:2013}, which is reviewed below. In this effective Lagrangian approach built from all spin 0 and non-derivative multi-quark interactions relevant at the scale of spontaneous chiral symmetry breaking, the complete set of interactions which break explicitely the chiral symmetry was included for the first time. This Lagrangian represents a generalization of the original Nambu-Jona-Lasinio \cite{Nambu:1961,Vaks} model extended to the realistic three flavor and color
case with $U(1)_A$ breaking six-quark 't Hooft interactions \cite{Hooft:1976,Hooft:1978,Bernard:1988,Bernard:1988a,Reinhardt:1988,Weise:1990,Vogl:1990,Weise:1991,Takizawa:1990,Klevansky:1992,Hatsuda:1994,Bernard:1993,Dmitrasinovic:1990,Birse:1996,Naito:2003} and an appropriate set of eight-quark interactions \cite{Osipov:2005b}. The last ones complete the number of vertices which are important in
four dimensions for dynamical $SU(3)_L\times SU(3)_R$ chiral symmetry breaking
\cite{Andrianov:1993a,Andrianov:1993b}. The Lagrangian considers all interactions relevant at the same order in large $N_c$ counting as the $U(1)_A$ anomaly term. 

The role of the new interactions contained in the explicit symmetry breaking (ESB) vertices has been analyzed at meson tree level approximation and in the isospin limit in connection with the low lying characteristics of the pseudoscalar and scalar meson nonets  \cite{Osipov1:2013},\cite{Osipov2:2013} and in the $T-\mu$ phase diagram associated with chiral transitions \cite{Moreira:2015}. An unprecedented accuracy for the description of the spectra has been achieved. One should stress that the present Lagrangian is able to account properly for the $SU(3)$ breaking effects in the description of the weak decay constants $f_\pi$ and $f_K$, in addition to yield the correct empirical $\eta$, $\eta'$ and $K$ meson masses, as well as the anomalous two photon decays of $\pi, \eta,\eta'$, in an unified description, which was an open problem for model versions without the ESB terms. 

The paper is organized as follows. In the next section is presented the effective multiquark Lagrangian and its bosonized form, the associated $N_c$ counting is reviewed,  in section III we address the mixing in the $\pi-\eta-\eta'$ system, the choices of representation of states, the decay parameters in the flavor basis,  and the compliance of the model in the approximation considered with the decay parameters transforming as the states. In section IV we present and discuss the numerical fits of the mass spectra and decay parameters. We end with a summary of the main results.
\vspace{0.5cm}

{\bf II. Survey of the model Lagrangian}

{\bf II a. Multiquark picture}
\vspace{0.5cm}

%%%%%%%%%%%%%%%%%%%%%%%%%%%%  TABLE 1  %%%%%%%%%%%%%%%%%%%%%%%%%%%%%%%%%%%%
%\vspace{0.5cm}
%\noindent
\begin{table*}
\label{pseudo}
\caption{The pseudoscalar  masses and weak decay constans (all in MeV) in the isospin limit used as input (marked with *) for different sets of the model. Parameter sets (a),(b)  contain explicit symmetry breaking interactions (see Table \ref{explicitSB}) and allow for a fit of the scalar masses and strong decays as well, $m_\sigma=550$ MeV, $m_\kappa=850$ MeV, $m_{a_0}=m_{f_0}=980$ MeV \cite{Osipov2:2013}; set (c) does not. Set (a) corresponds to an octet-singlet mixing angle in the scalar sector of $\theta_S=27.5^\circ$, set (b) to $\theta_S=25^\circ$.}
\label{table-1}
\begin{tabular*}{\textwidth}{@{\extracolsep{\fill}}lrrrrrrl@{}}
\hline
 Sets    & \multicolumn{1}{c}{$m_\pi$}
     & \multicolumn{1}{c}{$m_K$}
     & \multicolumn{1}{c}{$m_\eta$}
     & \multicolumn{1}{c}{$m_{\eta'}$}
     & \multicolumn{1}{c}{$f_\pi$}
     & \multicolumn{1}{c}{$f_K$} \\
\hline
a    & 138* & 494* & 547* & 958*  & 92*  & 113*    \\
b    & 138* & 494* & 547* & 958*  & 92*  & 113*    \\
c    & 138* & 494* & 475* & 958*  & 92*  & 115.7    \\
\hline
\end{tabular*}
\end{table*}
%%%%%%%%%%%%%%%%%%%%%%%%%%%%%%%%%%%%%%%%%%%%%%%%%%%%%%%%%%%%%%%%%%%%%%%%%%%
%%%%%%%%%%%%%%%%%%%%%%%%%%%%%%%%%%%%%%%%%%%%%%%%%%%%%%%%%%%%%%%
%\vspace{0.5cm}
%\noindent
\begin{table*}
\caption{Parameter sets of the model: $m_u=m_d={\hat m}, m_s$, and $\Lambda$ are given
         in MeV. The couplings have the following units: $[G]=$ GeV$^{-2}$,
         $[\kappa ]=$ GeV$^{-5}$, $[g_1]=[g_2]=$ GeV$^{-8}$. We also show here
         the values of constituent quark masses $M_u=M_d={\hat M}$ and $M_s$ in MeV.
         See also caption of Table \ref{table-1}.}
\label{table-2}
\begin{tabular*}{\textwidth}{@{\extracolsep{\fill}}lrrrrrrrrl@{}}
\hline
Sets & \multicolumn{1}{c}{${\hat m}$}
     & \multicolumn{1}{c}{$m_s$}
     & \multicolumn{1}{c}{${\hat M}$}
     & \multicolumn{1}{c}{$M_s$}
     & \multicolumn{1}{c}{$\Lambda$}
     & \multicolumn{1}{c}{$G$}
     & \multicolumn{1}{c}{$-\kappa$}
     & \multicolumn{1}{c}{$g_1$}
     & \multicolumn{1}{c}{$g_2$} \\
\hline
a  & 4.0* & 100* & 373 & 544 & 828  & 10.48  & 122.   & 3284  &173  \\
b  & 4.0* & 100* & 372 & 542 & 829  & 9.83  & 118.5   & 3305  &-158 \\
c  & 6.1 & 190 & 375 & 569 & 836  &9.79  & 138.2   & 2500* &100*  \\
\hline
\end{tabular*}
\end{table*}
%%%%%%%%%%%%%%%%%%%%%%%%%%%%%%%%%%%%%%%%%%%%%%%%%%%%%%%%%%%%%%%%
%\vspace{0.5cm}
%\noindent
\begin{table*}
\caption{Explicit symmetry breaking interaction couplings. The couplings have
the following units:   $[\kappa_2]=$ GeV$^{-3}$,
$[g_3]=[g_4]=$ GeV$^{-6}$, $[g_5]=[g_6]=[g_7]=[g_8]=$ GeV$^{-4}$. See also caption of Table \ref{table-1}.}
\label{explicitSB}
\begin{tabular*}{\textwidth}{@{\extracolsep{\fill}}lrrrrrrrrrl@{}}
\hline
Sets  & \multicolumn{1}{c}{$\kappa_2$}
      & \multicolumn{1}{c}{$-g_3$}
      & \multicolumn{1}{c}{$g_4$}
      & \multicolumn{1}{c}{$g_5$}
      & \multicolumn{1}{c}{$-g_6$}
      & \multicolumn{1}{c}{$-g_7$}
      & \multicolumn{1}{c}{$g_8$} \\
\hline
a   & 6.17  & 6497 & 1235& 213 & 1642  & 13.3 & -64   \\
b   & 5.61  & 6472 & 702 & 210 & 1668  & 100  & -38   \\
c   & 0  & 0 & 0 & 0 & 0  & 0  & 0   \\  
\hline
\end{tabular*}
\end{table*}
%%%%%%%%%%%%%%%%%%%%%%%%%%%%%%%%%%%%%%%%%%%%%%%%%%%%%%%%%%%%%%%%%%%%%%%%%%%%%%
\begin{table*}
\caption{The mixing angles in the $\eta-\eta'$ system in isospin limit, and related weak decay constants for the sets discussed and in comparison with different approaches, see also main text (the systematic error estimates given in \cite{Escribano:2015},\cite{Babar} have been omitted here ).}
\label{0-8angles}
\begin{tabular*}{\textwidth}{@{\extracolsep{\fill}}lrrrrrrrrrrrl@{}}
\hline
 Sets & \multicolumn{1}{c}{$\theta_P^\circ$}
      & \multicolumn{1}{c}{$\theta_0^\circ$}
      & \multicolumn{1}{c}{$\theta_8^\circ$}
      & \multicolumn{1}{c}{$\frac{f_0}{f_\pi}$}
      & \multicolumn{1}{c}{$\frac{f_8}{f_\pi}$} \\
\hline
a    & -12* &-1.42 & -21.37 & 1.172 &1.318   \\
b    & -15* &-4.42 & -24.37 & 1.172 &1.322   \\
c    & -14.5 &-2.82 & -24.78 & 1.197 &1.365   \\
\cite{Feldmann:1998}  phen.   & -13.3 & -6.8 & -19.4 & 1.10 & 1.19    \\
\cite{Feldmann:1998} phen.    & -15.4 & -9.2 & -21.2 & 1.17 & 1.26    \\
\cite{Goity:2002} CHPT & -10.5 & -1.5 & -20.0 & 1.24 & 1.31    \\
\cite{Kaiser:2000} CHPT & - & -4. & -20.5 & 1.10 & 1.28    \\
\cite{de Fazio:2000} sum rules & - & -15.6 & -10.8 & 1.39 & 1.39    \\
\cite{Escribano:2015} Pad\'e approximants $(\eta)$ & -16.4 & -11.3 & -21.3 & 1.15 & 1.22\\
\cite{Escribano:2015} Pad\'e approximants $(\eta')$ & -13.3 & -1.5 & -24.2 & 1.28 & 1.46\\
\cite{Babar} BABAR (from \cite{Escribano:2015} $(\eta)$) & -21.7 & -26.7 & -16.5& 1.04 &0.98 \\
\cite{Babar} BABAR (from \cite{Escribano:2015} $(\eta')$) & -17.7 & -15.6 & -19.9& 1.14 &1.11 \\ 
\hline
\end{tabular*}
\end{table*}
%%%%%%%%%%%%%%%%%%%%%%%%%%%%%%%%%%%%%%%%%%%%%%%%%%%%%%%%%%%%%%%
%\vspace{0.5cm}
%\noindent
\begin{table*}
\caption{Empirical input used in the fits with isospin breaking, sets A and B with ESB interactions, set C without. Primes indicate which masses of the pion and kaon  multiplets have been used for the fit, the other being output. Masses in units of MeV, angle $\psi$ in degrees.}
\label{mi}
\begin{tabular*}{\textwidth}{@{\extracolsep{\fill}}lrrrrrrrrrl@{}}
\hline
Sets & \multicolumn{1}{c}{${m_\pi^0}$}
     & \multicolumn{1}{c}{${m_\pi^{\pm}}$}
     & \multicolumn{1}{c}{$m_\eta$}
     & \multicolumn{1}{c}{${m_\eta'}$}
     & \multicolumn{1}{c}{$m_K^0$}
     & \multicolumn{1}{c}{$m_K^{\pm}$}
     & \multicolumn{1}{c}{$f_\pi$}
     & \multicolumn{1}{c}{$f_K$}
     & \multicolumn{1}{c}{$\psi$} \\
\hline
A,B  & 136' & 136.6 & 547 & 958 & 500 &494' & 92  & 113  & 39.7  \\
C    & 136' & 137.0 & 477 & 958 & 501& 497'  & 92  & 116   & 39.7   \\
%C & 136& 136.6 & 477 & 958 &   500& 497  & 92  & 116   & -15   \\
\hline
\end{tabular*}
\end{table*}
%%%%%%%%%%%%%%%%%%%%%%%%%%%%%%%%%%%%%%%%%%%%%%%%%%%%%%%%%%%%%%%%

%%%%%%%%%%%%%%%%%%%%%%%%%%%%%%%%%%%%%%%%%%%%%%%%%%%%%%%%%%%%%%%%%%%%%%%%%%%
%%%%%%%%%%%%%%%%%%%%%%%%%%%%%%%%%%%%%%%%%%%%%%%%%%%%%%%%%%%%%%%
%\vspace{0.5cm}
%\noindent
\begin{table*}
\caption{Parameter sets with isospin breaking, set A with ESB interactions, set B without.  The couplings have the following units: $[G]=$ GeV$^{-2}$,
         $[\kappa ]=$ GeV$^{-5}$, $[g_1]=[g_2]=$ GeV$^{-8}$, $[\kappa_2]=$ GeV$^{-3}$,
$[g_3]=[g_4]=$ GeV$^{-6}$, $[g_5]=[g_6]=[g_7]=[g_8]=$ GeV$^{-4}$. $\Lambda$ is given
         in MeV. 
         See also caption of Table \ref{table-1}.}
\label{ibpar}
\begin{tabular*}{\textwidth}{@{\extracolsep{\fill}}lrrrrrrrrrrrl@{}}
\hline
Sets & \multicolumn{1}{c}{${G}$}
     & \multicolumn{1}{c}{$-\kappa$}
     & \multicolumn{1}{c}{${g_1}$}
     & \multicolumn{1}{c}{$g_2$}
     & \multicolumn{1}{c}{$\Lambda$}
     & \multicolumn{1}{c}{$\kappa_2$}
     & \multicolumn{1}{c}{$g_3$}
     & \multicolumn{1}{c}{$g_4$}
     & \multicolumn{1}{c}{$g_5$} 
     & \multicolumn{1}{c}{$g_6$}
     & \multicolumn{1}{c}{$g_7$}
     & \multicolumn{1}{c}{$g_8$} \\
\hline
A  & 10.48 & 116.8 & 3284 & 1237 & 828.5  & 6.24  & 2365   & 1182  &160 & 712 & 580 & 44  \\
B  & 10.48 & 116.8 & 3284 & 1252 & 828.5  & 6.26  & 2481   & 1182  &151 & 745 & 591 & 49  \\
C  & 9.79 & 137.4 & 2500 &   117  & 835.7  & 0   & 0  &0 &0 & 0& 0 & 0 \\
\hline
\end{tabular*}
\end{table*}
%%%%%%%%%%%%%%%%%%%%%%%%%%%%%%%%%%%%%%%%%%%%%%%%%%%%%%%%%%%%%%%%

%\vspace{0.5cm}
%\noindent
\begin{table*}
\caption{Isospin breaking parameter $r=\frac{m_d-m_u}{m_d+m_u}$, current and constituent quark masses $m_u$, $m_d$,$m_s$ $M_u$, $M_d$, $M_s$ in MeV and $\pi^0-\eta$, $\pi^0-\eta'$ mixing angles $\epsilon$ and $\epsilon'$. }
\label{isobreak}
\begin{tabular*}{\textwidth}{@{\extracolsep{\fill}}lrrrrrrrrrrl@{}}%
%\begin{tabular*}{\textwidth}{@{\extracolsep{\fill}}lrrl@{}}
\hline
Sets  & \multicolumn{1}{c}{$r$}
      & \multicolumn{1}{c}{$m_u$}
      & \multicolumn{1}{c}{$m_d$}
      & \multicolumn{1}{c}{$m_s$}
      & \multicolumn{1}{c}{$M_u$}
      & \multicolumn{1}{c}{$M_d$}
      & \multicolumn{1}{c}{$M_s$}
     % & \multicolumn{1}{c}{$m_{\pi^0}$}
      & \multicolumn{1}{c}{$\epsilon$}
      & \multicolumn{1}{c}{$\epsilon'$} 
      & \multicolumn{1}{c}{$\frac{\epsilon}{\epsilon'}$} \\
\hline
A  & 0.372* & 2.179   & 4.760 & 95*& 372  & 375 &544& 0.014*  &  0.0037*  & 3.78   \\
B  & 0.372* & 2.166   & 4.733 & 95*& 372& 375 &544 & 0.017*  &  0.0045*  & 3.95   \\
C  & 0.372* & 3.774   & 8.246 & 194 & 373  & 380 &573  & 0.022  & 0.0025  & 8.78   \\
%C  & 0.325*  & 4.043  & 7.936 &194 & 374  & 380 &573  & 0.0194  & 0.0022  & 8.78   \\
%a  & 0.25  & 3.0  & 5.0 & 368.4  & 348.4 &137.0  & 0.021  & 0.0039  & 5.47   \\
%a'  & 0.55  & 1.8  & 6.2 & 370.0  & 377.0 &137.5  & 0.015  & 0.0027  & 5.45   \\  
%b  & 0.175 & 3.3 &  4.7 & 368.6 & 375.6  &137.5  & 0.0157  & 0.0026 & 6.08    \\  
%b  & 0.20  & 3.2  & 4.8 & 368.1 & 376.1 &137.3  &  0.0180  & 0.00296 & 6.08    \\  
%b  & 0.25  & 3.0 & 5.0 & 367.1 & 377.1 &137.0  & 0.0225  & 0.0037 & 6.08    \\ 
%c  & 0.175 & 5.0 &  7.2 & 373.7 & 377.1  &137.8  & 0.0107  & 0.00123 & 8.75    \\  
%c  & 0.20  & 4.9  & 7.3 & 373.4 & 377.3 &137.8  &  0.0123  & 0.00140 & 8.75    \\  
%c  & 0.25  & 4.6 & 7.6 & 372.9 & 377.8 &137.6  & 0.0153  & 0.00175 & 8.75    \\   
\hline
\end{tabular*}
\end{table*}
\begin{table*}
\caption{$\epsilon$ and $\epsilon'$ values in the literature. }
\label{epp}
\begin{tabular*}{\textwidth}{@{\extracolsep{\fill}}lrrrrrrrrl@{}}
\hline
      & \multicolumn{1}{c}{$\epsilon$}
      & \multicolumn{1}{c}{$\epsilon'$}
      & \multicolumn{1}{c}{$\frac{\epsilon}{\epsilon'}$} \\
     % & \multicolumn{1}{c}{$-z$} \\
\hline
\cite{Feldmann:1999} phen.  & 0.014 & 0.0037 & 3.78     \\
\cite{Kroll:2005} phen.  & $0.017\pm 0.002$ & $0.004\pm 0.001$ & $4.25\pm 1.17$    \\
\cite{Goity:2002} ChPt NLO & $0.014\div 0.016$ & - & -    \\  
\cite{Coon:1986} phen. & 0.021 & - & -   \\  
\cite{BES:2004} Exp.  & $0.030 \pm 0.002$ & - &- \\  
\cite{Tippens:2001} Exp.  & $0.026\pm 0.007$ & - & -   \\ 
\hline
\end{tabular*}
\end{table*}
%%%%%%%%%%%%%%%%%%%%%%%%%%%%%%%%%%%%%%%%%%%%%%%%%%%%%%%%%%%%%%%%%%%%%%%%%%%%
The Langrangian considered is built from all spin 0 chiral $SU(3)_L \times SU(3)_R$ symmetric and parity conserving combinations relevant at the scale $\Lambda$ of spontaneous breaking of chiral symmetry. This means that in the corresponding effective potential are kept all the interactions  which do not vanish in the $\Lambda \rightarrow \infty$ limit.
These consist of vertices involving non-derivative  quark-antiquark fields, denoted by $\Sigma$, with $\Sigma=\frac{1}{2}(s_a-i p_a)\lambda_a$ and $s_a=\bar q\lambda_aq,\, p_a=\bar qi\gamma_5
\lambda_aq$, $\lambda_a$ being the standard Gell-Mann matrices for $a=1...8$ and $\lambda_0= \sqrt{2/3}\times 1$, as well as of their interactions with the external sources $\chi$ which transform as $\chi=(3,3*)$ under $SU(3)_L \times SU(3)_R$, 
\begin{equation}
\label{LQ}
   L=\bar qi\gamma^\mu\partial_\mu q+L_{int}+L_\chi.
\end{equation}
The term  
%Let us consider now the structure of multi-quark vertices in detail
%\cite{Osipov:2013}. The Lagrangian corresponding to the case (i) is well known
\begin{eqnarray}
\label{L-int}
   L_{int}&=&\frac{\bar G}{\Lambda^2}\mbox{tr}\left(\Sigma^\dagger\Sigma\right)
   +\frac{\bar\kappa}{\Lambda^5}\left(\det\Sigma+\det\Sigma^\dagger\right)
   \nonumber \\
   &+&\frac{\bar g_1}{\Lambda^8}\left(\mbox{tr}\,\Sigma^\dagger\Sigma\right)^2
   +\frac{\bar g_2}{\Lambda^8}\mbox{tr}
   \left(\Sigma^\dagger\Sigma\Sigma^\dagger\Sigma\right).
\end{eqnarray}
is well known. Here and elsewhere we use barred quantities for any dimensionless coupling, these are related to the dimensionful ones through powers of the scale $\Lambda$,  $g_i={\bar g}_i/\Lambda^\gamma$. $L_{int}$ contains the leading order (LO) in $N_c$ four quark (q) NJL interactions with coupling $\bar G$, generalized to the 3 flavor case, the NLO 6q 't Hooft $U(1)_A$ breaking flavor determinant with coupling $\bar \kappa$, and the two possible 8q interactions $\sim {\bar g_1},{\bar g_2}$, which have the same $N_c$ counting as the 't Hooft term.  We refer to  \cite{Osipov1:2013},\cite{Osipov2:2013} for a detailed discussion of the large $N_c$ counting scheme which complies with the counting rules based on powers of the scale $\Lambda$ of spontaneous breaking of chiral symmetry.  Rephrasing it, means that the terms which vanish as $\Lambda \rightarrow \infty$ are the same that also vanish on grounds of $N_c\rightarrow \infty$. Applying these rules, summarized after eq. {\ref{L-chi-1}, the following set of
terms involving interactions with the sources $\chi$ emerge, which act also up to the same order in $N_c$ counting as the 't Hooft term
\begin{equation}
   L_\chi =\sum_{i=0}^{10}L_i,      
\end{equation}
where
\begin{eqnarray}
\label{L-chi-1}
   L_0&=&-\mbox{tr}\left(\Sigma^\dagger\chi +\chi^\dagger\Sigma\right)
   \nonumber \\
   L_1&=&-\frac{\bar\kappa_1}{\Lambda}e_{ijk}e_{mnl}
   \Sigma_{im}\chi_{jn}\chi_{kl}+h.c.
   \nonumber \\
   L_2&=&\frac{\bar\kappa_2}{\Lambda^3}e_{ijk}e_{mnl}
   \chi_{im}\Sigma_{jn}\Sigma_{kl}+h.c.
   \nonumber \\
   L_3&=&\frac{\bar g_3}{\Lambda^6}\mbox{tr}
   \left(\Sigma^\dagger\Sigma\Sigma^\dagger\chi\right)+h.c.
   \nonumber \\
   L_4&=&\frac{\bar g_4}{\Lambda^6}\mbox{tr}\left(\Sigma^\dagger\Sigma\right)
   \mbox{tr}\left(\Sigma^\dagger\chi\right)+h.c.
   \nonumber \\
   L_5&=&\frac{\bar g_5}{\Lambda^4}\mbox{tr}\left(\Sigma^\dagger\chi
   \Sigma^\dagger\chi\right)+h.c.
   \nonumber \\
   L_6&=&\frac{\bar g_6}{\Lambda^4}\mbox{tr}\left(\Sigma\Sigma^\dagger\chi
   \chi^\dagger +\Sigma^\dagger\Sigma\chi^\dagger\chi\right)
   \nonumber \\
   L_7&=&\frac{\bar g_7}{\Lambda^4}\left(\mbox{tr}\Sigma^\dagger\chi
   + h.c.\right)^2
   \nonumber \\
   L_8&=&\frac{\bar g_8}{\Lambda^4}\left(\mbox{tr}\Sigma^\dagger\chi
   - h.c.\right)^2
   \nonumber \\
   L_9&=&-\frac{\bar g_9}{\Lambda^2}\mbox{tr}\left(\Sigma^\dagger\chi
   \chi^\dagger\chi\right)+h.c.
   \nonumber \\
   L_{10}&=&-\frac{\bar g_{10}}{\Lambda^2}\mbox{tr}\left(\chi^\dagger\chi\right)
   \mbox{tr}\left(\chi^\dagger\Sigma\right)+h.c.
\end{eqnarray}
We recall that the $N_c$ book keeping is as follows: 
%Some useful insight into the Lagrangian above can be obtained by considering it
%from the point of view of the $1/N_c$ expansion. Indeed, the number of color 
%components of the quark field $q^i$ is $N_c$, hence summing over color indices 
%in $\Sigma$ gives a factor of $N_c$, i.e. one counts 

\begin{eqnarray}
\label{cr}
&&\Sigma\sim N_c, \quad \Lambda\sim N_c^0=1, \quad  G\sim 1/N_c, \quad \kappa\sim 1/N_c^3,  \nonumber \\
&&\kappa_1, g_9, g_{10}\sim 1/N_c,  \nonumber \\
&&\kappa_2, g_5, g_6, g_7, g_8\sim 1/N_c^2, \quad g_3, g_4\sim 1/N_c^3. 
\end{eqnarray}
     
The scale $\Lambda$ which gives the right dimensionality to the multiquark vertices is related with the cutoff on the divergent quark one-loop integrals, its $N_c$ counting is  dictated by the gap equations in the phase of spontaneously broken chiral symmetry,  eqs. (\ref{ji}),(\ref{PVk}),({\ref{gap}) below.

The terms $L_1 ... L_{10}$ of Lagrangian (\ref{L-chi-1}) have been introduced recently in the model  \cite{Osipov1:2013},\cite{Osipov2:2013}  and generalize to NLO in $N_c$ the explicit symmetry breaking (ESB) standard LO mass term $L_0$.
%The cut-off $\Lambda$ that gives the right dimensionality to the multi-quark 
%vertices scales as $\Lambda\sim N_c^0=1$. On the other hand, since the leading 
%quark contribution to the vacuum energy is known to be of order $N_c$, the 
%first term in (\ref{L-int}) is estimated as $N_c$, and we conclude that $G\sim 
%1/N_c$. 

%Furthermore, the $U(1)_A$ anomaly contribution (the second term in 
%(\ref{L-int})) is suppressed by one power of $1/N_c$, it yields $\kappa\sim 
%1/N_c^3$. 

%The last two terms in (\ref{L-int}) have the same $N_c$ counting as the 
%Hooft term. They are of order $1$. Indeed, Zweig's rule violating effects
%are always of order $1/N_c$ with respect to the leading order contribution 
%$\sim N_c$. This reasoning helps us to find $g_1\sim 1/N_c^4$. The term with 
%$g_2\sim 1/N_c^4$ is also $1/N_c$ suppressed. It represents the next to the 
%leading order contribution with one internal quark loop in $N_c$ counting. 
%Such vertex contains the admixture of the four-quark component $\bar qq\bar qq$
%to the leading quark-antiquark structure at $N_c\to\infty$.  

%Next, all terms in eq. (\ref{L-chi-1}), except $L_0$, are of order 1. The 
%argument is just the same as before: this part of the Lagrangian is obtained 
%by succesive insertions of the $\chi$-field ($\chi$ counts as $\chi\sim 1$) in 
%means that $\kappa_1, g_9, g_{10}\sim 1/N_c$, $\kappa_2, g_5, g_6, g_7, 
%g_8\sim 1/N_c^2$, and $g_3, g_4\sim 1/N_c^3$.  
   
Since all the blocks $L_0...L_{10}$  conform with the symmetry pattern of the model one is free to  choose for the source the constant valued matrix $\chi= {\cal M}/2$, 
\begin{equation}
\label{mcur}
{\cal M}= \mu_a \lambda_a= diag(\mu_u,\mu_d,\mu_s).
\end{equation}
Whenever convenient we use in the following the equivalent redefinition of flavor indices from $a=0,3,8$ to $i=u,d,s$ for any observable $A$  \cite{Osipov:2002}
\begin{equation}
\label{indices}
   A_a =e_{a i} A_i, \quad
   e_{a i}=\frac{1}{2\sqrt 3}\left(
          \begin{array}{ccc}
          \sqrt 2&\sqrt 2&\sqrt 2 \\
          \sqrt 3&-\sqrt 3& 0 \\
          1&1&-2
          \end{array} \right).
\end{equation}
%the standard identification for the non-vanishing components $\mu_0= \sqrt{1/6} (\mu_u + \mu_d + \mu_s), \mu_3 = (\mu_u - \mu_d)/2, \mu_8 = \sqrt{3} (\mu_u + \mu_d - 2 \mu_s)/6)$.

The terms $L_1,L_9,L_{10}$ are related to the Kaplan-Manohar ambiguity \cite{Manohar:1986,Leutwyler:1990,Donoghue:1992,Leutwyler:1996} in the definition of the quark mass. In the present case this corresponds to the following freedom in transforming the external source \cite{Osipov1:2013}
\begin{eqnarray}
\label{ci}
   \chi^{(c_i)}&=&\chi +\frac{c_1}{\Lambda}
   \left(\det\chi^\dagger\right)\chi\left(\chi^\dagger\chi\right)^{-1}+
   \frac{c_2}{\Lambda^2}\chi\chi^\dagger\chi  
   \nonumber \\
   &+&\frac{c_3}{\Lambda^2}\mbox{tr}\left(\chi^\dagger\chi\right)\chi
\end{eqnarray} 
with three independent constants $c_i$, which has the same symmetry transformation
property as $\chi$.  The description in terms of  $\chi^{(c_i)}$ in the place of $\chi$ is equivalent, leading to the same Lagrangian, with some of the couplings redefined as 
\begin{eqnarray}
\label{rt}
   &&\bar\kappa_1\to\bar\kappa_1'=\bar\kappa_1+\frac{c_1}{2}, \quad 
   \bar g_5\to\bar g_5'=\bar g_5-\bar\kappa_2c_1,
   \nonumber \\
   &&\bar g_7\to\bar g_7'=\bar g_7+\frac{\bar\kappa_2}{2}c_1,  \quad   
   \bar g_8\to\bar g_8'=\bar g_8 +\frac{\bar\kappa_2}{2}c_1, \quad
   \nonumber \\
   &&\bar g_9\to\bar g_9'=\bar g_9 +c_2 -2\bar\kappa_1 c_1, \quad 
   \nonumber \\
   &&\bar g_{10}\to\bar g_{10}'=\bar g_{10}+c_3+2\bar\kappa_1 c_1.    
\end{eqnarray}
Note that the redefinition of the $\chi$ fields (\ref{ci}) in the model contains more terms than the one usually considered, in the context of second order ChPT, which leads to the following redefinition of the current quark mass \cite{Manohar:1986}
\begin{equation}
{\cal M}\rightarrow {\cal M}(\lambda)= {\cal M}+\lambda {\cal M} ({\cal M}^\dagger {\cal M})^{-1} det {\cal M}^\dagger.
\end{equation}

It has been reported that the Kaplan-Manohar ambiguity may be in conflict with the large $N_c$ counting rules of ChPT \cite{Kaiser:2000}.  Following from a threefold chiral expansion in the number of derivatives,  powers of quark masses and 
powers of $1/N_c$, as well as the counting associated with the $\theta$ angle related with the $U_A(1)$ sector, $\lambda$ is found to be suppressed to all orders in the large $N_c$ limit \cite{Kaiser:2000}. 
%\begin{equation}
%M\rightarrow M(\lambda)= M+\lambda M (M^\dagger M)^{-1} det M^\dagger]
%\end{equation}

Regarding our model Lagrangian,  the new couplings, the primed ones in  (\ref{rt}), must not dominate over the $N_c$ dependence of the unprimed ones.  This means that  in their leading contributions they scale at most as the unprimed couplings, $c_i \sim 1/N_c$, see (\ref{cr}) . Attributing this counting to the $c_i$   suffices to warrant  that one can use the reparametrization freedom (\ref{rt}), in particular to obtain  ${\bar\kappa}_1' =\bar g_9' =\bar g_{10}'=0$. Indeed, remembering that our model Lagrangian  has been constructed to incorporate the classes of multiquark interactions which do not vanish in the effective potential as $N_c\rightarrow \infty$ in the phase of spontaneously broken chiral symmetry,  one gets for this  case that  ${\bar \kappa}_1=- c_1/2$, and that up to subleading corrections (not considered at this order of the effective potential),   $\bar g_9=-c_2, {\bar g}_{10}=-c_3$,  ${\bar g'}_5={\bar g}_5,{\bar g'}_7={\bar g}_7, {\bar g'}_8={\bar g}_8$.

\vspace{0.5cm}

{\bf II b. Bosonized version}
\vspace{0.5cm}

The low energy meson characteristica are obtained after path integral bosonization of the quark Lagrangian
(\ref{LQ}). Following \cite{Reinhardt:1988} one may equivalently
use the introduced $s_a,p_a$ as auxiliary fields, and a further set of physical scalar and pseudoscalar fields $\sigma =\sigma_a
\lambda_a,\,\phi = \phi_a\lambda_a$ to obtain the vacuum persistence amplitude of the theory as
\begin{eqnarray}
\label{genf3}
   Z&\!\!\! =\!\!\!&\int\!\! {\cal D}q{\cal D}\bar{q}
     \prod_a{\cal D}\sigma_a\prod_a{\cal D}\phi_a\
     \exp\left(i\!\!\int\!\!\ud^4x
     {L}_q(\bar{q},q,\sigma ,\phi )\right)
     \nonumber \\
    &\!\!\!\times\!\!\!& \int
     \limits^{+\infty }_{-\infty }\!
     \prod_a{\cal D}  s_a 
     \prod_a{\cal D}p_a \
     \exp\left(i\!\!\int\!\!\ud^4x{ L}_{aux}(\sigma ,\phi
     ; s,p )\right). \nonumber \\
\end{eqnarray}
In these variables the Lagrangian reads
\begin{eqnarray}
\label{Lparts}
&&L=L_q +L_{aux} \nonumber \\
&&L_q=\bar q\left(i\gamma^\mu\partial_\mu -(\sigma + M) - i\gamma_5\phi
   \right)q \nonumber \\
&&L_{aux}\!=\!s_a (\sigma_a + M_a-m_a) + p_a\phi_a +L_{int}(s,p) \nonumber \\
   &\!+\!&\sum_{i=2}^8 L_i'(s,p,m).
\end{eqnarray}
Here $L_i'$ contains the ESB terms  and $L_0$ appears as $s_am_a$ in $L_{aux}$. The external scalar fields $\sigma$ have been shifted to $\sigma \rightarrow \sigma + M$, so that the expectation value of the shifted
fields in the vacuum corresponding to dynamically broken chiral symmetry
vanish. The expectation value of the �unshifted� scalar field
$<\sigma> = M_a\lambda_a = diag(M_u,M_d,M_s)$ corresponds to the point where the effective potential of the theory achieves its minimum, with $M$ being the constituent quark masses. 
In (\ref{Lparts}) the bilinear form in the quark fields  $L_q$ can be integrated out from the path integral, and results in the fermion determinant (see (\ref{W}) below), which generates the kinetic terms of the $\sigma,\phi$ fields. The remaining integrations are over a static non-Gaussian system of  $s,p$ fields, and are done in the stationary phase approximation (SPA). First we present the results for the SPA integration, obtained
from the extremum conditions
\begin{equation}
\label{sp}
   \frac{\partial L}{\partial s_a}=0, \quad \frac{\partial L}{\partial p_a}=0,
\end{equation}
which must be fulfilled in the neighbourhood of the uniform vacuum state of the
theory.  The solutions of eq.
(\ref{sp}) are seeked in the form:
\begin{eqnarray}
\label{st}
   s_a^{st}&=&h_a+ h_{ab}^{(1)}\sigma_b + h_{abc}^{(1)}\sigma_b\sigma_c
   +h_{abc}^{(2)}\phi_b\phi_c + \ldots
   \nonumber \\
    p_a^{st}&=&h_{ab}^{(2)}\phi_b + h_{abc}^{(3)}\phi_b\sigma_c +\ldots
\end{eqnarray}
Eqs. (\ref{sp}) determine all coefficients of this expansion giving rise to a
system of cubic equations to obtain $h_a$, eq. (\ref{h}), and a full set of recurrence
relations to find higher order coefficients in (\ref{st}). The result is cast in the form
\begin{eqnarray}
\label{lam}
   L_{aux}\!\!\!\!\!\!
   &&=h_a\sigma_a+\frac{1}{2}\,h_{ab}^{(1)}\sigma_a\sigma_b
      +\frac{1}{2}\,h_{ab}^{(2)}\phi_a\phi_b  \\
   &&+\,\frac{1}{3}\,\sigma_a\left[h^{(1)}_{abc}\sigma_b\sigma_c 
  +\left(h^{(2)}_{abc}+h^{(3)}_{bca}\right)\phi_b\phi_c\right]+ \ldots \nonumber
\end{eqnarray}
Here $h_a$ are related to the quark condensates, $h_{ab}^{(1)}$,
$h_{ab}^{(2)}$ contribute to the masses of scalar and pseudoscalar states respectively, and higher indexed
$h$'s are the couplings that measure the strength of the
meson-meson interactions. %The transition from the Lagrangian $L_{aux}(s,p)$ in
%(\ref{L}) to its form $L_{aux}(\sigma,\phi)$ in (\ref{lam}) can be viewed as a
%Legendre transformation. 
%We proceed now to explain the details of determining $h$. We address first the
%coefficients $h_a$, $h_{ab}^{(1)}$, and $h_{ab}^{(2)}$. In particular, eq.
%(\ref{sp}) states that $h_a=0$, if $a\neq 0,3,8$, while $h_\alpha$ ($\alpha
%=0,3,8$), after the convenient redefinition to the flavor indices $i=u,d,s$
%\begin{equation}
%   h_\alpha =e_{\alpha i} h_i, \quad
%   e_{\alpha i}=\frac{1}{2\sqrt 3}\left(
%          \begin{array}{ccc}
%          \sqrt 2&\sqrt 2&\sqrt 2 \\
%          \sqrt 3&-\sqrt 3& 0 \\
%          1&1&-2
%          \end{array} \right),
%\end{equation}
From eq. (\ref{sp}) and using (\ref{indices}) one obtains the following system of cubic equations for the one index coefficients $h_i$, $(i=\{u,d,s\})$
\begin{eqnarray}
\label{h}
   &&M_i-m_i + \frac{\kappa}{4}t_{ijk}h_jh_k +\frac{h_i}{2}\left(
     2G+g_1h^2 + g_4 m h\right) + \frac{g_2}{2}h_i^3 \nonumber \\
   &&+\frac{m_i}{4}\left[3g_3h_i^2 +g_4h^2 +2(g_5+g_6)m_ih_i
     +4g_7 m h\right] \nonumber \\
   &&+\kappa_2t_{ijk}m_jh_k=0.
\end{eqnarray}
Here $t_{ijk}$ is a totally symmetric quantity, whose
nonzero components are $t_{uds}=1$; there is no summation over the open index
$i$ but we sum over the dummy indices, e.g. $h^2=h_u^2+h_d^2+h_s^2,
m h=m_uh_u+m_dh_d+m_sh_s$. 
Regarding the $h$ coefficients with more than one index, we indicate explicitly only the expression needed for the present study of the pseudoscalar masses (the complete expressions up to 3 indices can be found in  \cite{Osipov1:2013},\cite{Osipov2:2013}) 
%\begin{eqnarray}
%\label{h1}
%   &&-2\left(h_{ab}^{(1)}\right)^{-1}= \left(2G+g_1h^2+g_4\mu h\right)\delta_{ab}
%   +4g_1h_ah_b \nonumber \\
%   &&+3A_{abc}\left(\kappa h_c+2\kappa_2\mu_c\right)
%   +g_2h_{r}h_{c}\left(d_{abe}d_{cre}+2d_{ace}d_{bre}\right)
%   \nonumber \\
%   &&+g_3\mu_{r}h_{c}\left(d_{abe}d_{cre}+d_{ace}d_{bre}+d_{are}d_{bce}
%   \right) \nonumber \\
%   &&+2g_4\left(\mu_ah_b+\mu_bh_a\right)
%   +g_5\mu_r\mu_c\left(d_{are}d_{bce}-f_{are}f_{bce}\right)
%   \nonumber \\
%   &&+g_6\mu_r\mu_cd_{abe}d_{cre}+4g_7\mu_a\mu_b.
%\end{eqnarray}
\begin{eqnarray}
\label{h2}
   &&-2\left(h_{ab}^{(2)}\right)^{-1}= \left(2G+g_1h^2+g_4 m h\right)\delta_{ab}
   \nonumber \\
   &&-3A_{abc}\left(\kappa h_c+2\kappa_2 m_c\right)
   +g_2h_{r}h_{c}\left(d_{abe}d_{cre}+2f_{are}f_{bce}\right)
   \nonumber \\
   &&+g_3 m_{r}h_{c}\left(d_{abe}d_{cre}+f_{are}f_{bce}+f_{ace}f_{bre}
   \right) \nonumber \\
   &&-g_5 m_r m_c\left(d_{are}d_{bce}-f_{are}f_{bce}\right)
   \nonumber \\
   &&+g_6 m_r m_c d_{abe}d_{cre}-4g_8 m_a m_b.
\end{eqnarray}
which can be readily inverted.
These coefficients are totally defined in terms of $h_a$ and the parameters of
the model. %Eqs. (\ref{h1})-(\ref{h2}) can be easily converted into explicit
%formulae for $h_{ab}^{(i)}$, $(i=1,2)$.

%In particular, eq. (\ref{cqmm}) reads in this basis
%\begin{equation}
%\label{cqm-2}
%   m_i=\mu_i\left(1+\frac{g_9}{4}\mu_i^2+\frac{g_{10}}{4}\mu^2\right)
%   +\frac{\kappa_1}{2}t_{ijk}\mu_j\mu_k.
%\end{equation}
%For the set $g_9=g_{10}=\kappa_1=0$ the current quark mass $m_i$ coincides
%precisely  with the explicit symmetry breaking parameter $\mu_i$.
Now to the fermion determinant related to the integration over the fermion fields:  we expand it using a heat-kernel technique that takes appropriately into account the quark mass differences, being chiral covariant at each order of the expansion
\cite{Osipov:2001b,Osipov:2001,Osipov:2001a},
\begin{eqnarray}
\label{W}
   W[Y]
   &\!=\!&\ln |\det D|=-\frac{1}{2} \int_0^\infty \frac{dt}{t} \rho(t)
   \exp\left(-t D_E^\dagger D_E\right), \nonumber \\
   D_E^\dagger D_E
   &\!=\!&M^2 -\partial^2 +Y, \quad Y=i\gamma_\mu (\partial_\mu +i\gamma_5
   \partial_\mu \phi )
   \nonumber \\
   &\!+\!&\sigma^2+\{M,\sigma\}+\phi^2+i\gamma_5[\sigma+M,\phi ],
\end{eqnarray}
or
\begin{equation}
\label{W1}
   W[Y]=-\int \frac{d^4 x_E}{32\pi^2} \sum_{i=0}^\infty I_{i-1} \mbox{tr}[b_i]
\end{equation}
where $D_E$ stands for the Dirac operator in Euclidean space. We consider the
expansion up to the third modified Seeley-DeWitt coefficient $b_i$
\begin{eqnarray}
\label{bi}
   b_0&\!=\!&1, \quad b_1=-Y,
   \nonumber \\
   b_2&\!=\!&\frac{Y^2}{2}+\frac{\lambda_3}{2} \Delta_{ud} Y +
   \frac{\lambda_8}{2 \sqrt{3}}(\Delta_{us}+\Delta_{ds}) Y,
\end{eqnarray}
with $\Delta_{ij}=M_i^2-M_j^2$. This order of the expansion takes into account
the dominant contributions of the quark one-loop integrals $I_i$ $(i=0,1,
\ldots )$; these are the arithmetic average values $I_i=\frac{1}{3}
[J_i(M_u^2)+J_i(M_d^2)+J_i(M_s^2)]$ where
\begin{equation}
\label{ji}
   J_i(m^2)=\int\limits_0^\infty\frac{{\rm d}t}{t^{2-i}}\rho
   (t\Lambda^2) e^{-t m^2},
\end{equation}
with the Pauli-Villars regularization kernel \cite{Pauli,Bernard:1996}
\begin{equation}
\label{PVk}
   \rho (t\Lambda^2)=1-(1+t\Lambda^2)\exp (-t \Lambda^2).
\end{equation}
which is equivalent to the sharp 4D cutoff regularization for the scalar integrals considered.
%In the following we need therefore only to know two of them (the lowest order
%$\sim b_0$ contributes to the effective potential and is not needed in the
%present study)
%\begin{equation}
%\label{j0}
%   J_0(m^2)=\Lambda^2- m^2\ln\left(1+\frac{\Lambda^2}{m^2}\right),
%\end{equation}
%and
%\begin{equation}
%\label{j1}
%      J_1(m^2)=\ln\left(1+\frac{\Lambda^2}{m^2}\right)
%      -\frac{\Lambda^2}{\Lambda^2+m^2}\ .
%\end{equation}
Both terms proportional to $b_1$ and $b_2$ have contributions to the gap
equations and meson masses, but only $b_2$ contributes to the kinetic and
meson interaction terms. By excluding the $\sigma$ tadpole from the total
Lagrangian, one obtains the gap equations
\begin{equation}
\label{gap}
   h_i+\displaystyle\frac{N_c}{6\pi^2} M_i
   \left[3I_0-\left(3M_i^2-M^2\right) I_1 \right]=0.
\end{equation}
where $M^2=M_u^2+M_d^2+M_s^2$. 
We now see that (\ref{gap}) must be solved self-consistently with the SPA equations (\ref{h}).
\vspace{0.5cm}

{\bf III. Mixing in the $\pi^0,\eta,\eta'$ system}

{\bf III a. Choices of  representations}
\vspace{0.5cm}
 
Finally one is ready to combine the terms of the total Lagrangian $L$ that contribute to
the kinetic terms $L_{kin}$ and meson masses $L_{mass}$ 
%Combining all
%terms of the total Lagrangian $L=L_{kin}+ L_{mass}+L_{int}$ that contribute to
%the kinetic terms $L_{kin}$ and meson masses $L_{mass}$ one gets
\begin{eqnarray}
\label{mass}
     &&L_{kin}+L_{mass}  \nonumber \\
     &\!=\!&\frac{N_cI_1}{16\pi^2}\,\mbox{tr}\left[
           (\partial_\mu \sigma )^2+(\partial_\mu \phi )^2\right]
           +\frac{N_cI_0}{4\pi^2}(\sigma_a^2+\phi_a^2)
           \nonumber \\
     &\!-\!&\frac{N_cI_1}{12\pi^2}
           \left\{
           \left[2\left(M_u+M_d\right)^2-M_u M_d -M_s^2\right](\sigma_1^2
           +\sigma_2^2)
           \right.
           \nonumber \\
     &\!+\!&\left[2\left(M_u+M_s\right)^2-M_u M_s -M_d^2\right]
           \left(\sigma_4^2+\sigma_5^2\right)
           \nonumber \\
     &\!+\!&\left[2(M_d+M_s)^2-M_dM_s-M_u^2\right]
           \left(\sigma_6^2+\sigma_7^2\right)
           \nonumber \\
     &\!+\!&\frac{1}{2}\left[\sigma_u^2\left(8 M_u^2-M_d^2 -M_s^2\right)
           +\sigma_d^2\left(8 M_d^2-M_u^2 -M_s^2\right)
           \right.
           \nonumber \\
     &\!+\!&\left.\sigma_s^2\left(8 M_s^2-M_u^2 -M_d^2\right)\right]
           \nonumber \\
     &\!+\!&\frac{1}{2}\left[\phi_u^2\left(2 M_u^2-M_d^2 -M_s^2\right)
           +\phi_d^2\left(2 M_d^2-M_u^2 -M_s^2\right)
           \right.
           \nonumber \\
     &\!+\!&\left.\phi_s^2\left(2 M_s^2-M_u^2 -M_d^2\right)\right]
           \nonumber \\
     &\!+\!&\left[2 \left(M_u-M_d\right)^2+M_uM_d-M_s^2\right]
           \left(\phi_1^2+\phi_2^2\right)
           \nonumber \\
     &\!+\!&\left[2 \left(M_u-M_s\right)^2+M_u M_s -M_d^2\right]
           \left(\phi_4^2+\phi_5^2\right)
           \nonumber \\
     &\!+\!&\left.\left[2\left(M_d-M_s\right)^2+M_dM_s-M_u^2\right]
           \left(\phi_6^2+\phi_7^2\right)\right\}
           \nonumber \\
     &\!+\!&\frac{1}{2}\,h_{ab}^{(1)}\sigma_a\sigma_b
     +\frac{1}{2}\,h_{ab}^{(2)}\phi_a\phi_b.
\end{eqnarray}
The kinetic term requires a redefinition of the meson fields,
\begin{equation}
\label{ren}
   \sigma_a =g\sigma_a^R, \quad \phi_a =g\phi_a^R, \quad
   g^2=\frac{4\pi^2}{N_cI_1},
\end{equation}
to obtain the standard factor $1/4$. 
%In this process one obtains the kinetic terms 
%for the physical fields $\phi$ and $\sigma$  in the expression \ref{mass} below),
The flavor and charged pseudoscalar fields are related
through
\begin{eqnarray}
\label{psm}
   &&\frac{\lambda_a}{\sqrt{2}}\phi_a =
     \pmatrix{
             \frac{\phi_u}{\sqrt{2}} &\pi^+& K^+ \cr
             \pi^- &\frac{\phi_d}{\sqrt{2}}& K^0 \cr
             K^- &{\bar K}^0 &\frac{\phi_s}{\sqrt{2}} \cr
             }
\end{eqnarray}
%   \nonumber\\
%   &&\frac{\lambda_a}{\sqrt{2}}\sigma_a =
%     \pmatrix{
%             \frac{\sigma_u}{\sqrt{2}} &a_0^+& \kappa^+ \cr
%             a_0^-&\frac{\sigma_d}{\sqrt{2}}& \kappa^0  \cr
%             \kappa^-&{\bar \kappa}^0&\frac{\sigma_s}{\sqrt{2}}\cr
%            }
%\end{eqnarray}
%and in particular for the diagonal components
%\begin{eqnarray}
%\label{dico}
%     \phi_3&=&\pi^0 \nonumber \\
%     \phi_u&=&\phi_3+\frac{\sqrt{2}\phi_0 +\phi_8}{\sqrt{3}}
%            =\phi_3+\eta_{ns} \nonumber \\
%     \phi_d&=&-\phi_3+\frac{\sqrt{2} \phi_0 +\phi_8}{\sqrt{3}}
%            =-\phi_3+\eta_{ns} \nonumber \\
%     \phi_s&=&\sqrt{\frac{2}{3}}\phi_0-\frac{2 \phi_8}{\sqrt{3}}
%            =\sqrt{2}\eta_s
%\end{eqnarray}
and similarly for the scalar fields. 
In the following we concentrate on the diagonal components of (\ref{psm}), which according to eq. (\ref{indices}) induce mixing between the $0,3,8$ field components of (\ref{mass}), in general. Indicating also the result of the  transformations discussed below in (\ref{sns}),(\ref{V}), we arrive at the following useful relations among fields   
\begin{eqnarray}
\label{dico}
%     \phi_3&=&\pi^0 \nonumber \\
     \phi_u&=&\phi_3+\frac{\sqrt{2}\phi_0 +\phi_8}{\sqrt{3}} 
            =\phi_3+\eta_{ns} \nonumber \\
     \phi_d&=&-\phi_3+\frac{\sqrt{2} \phi_0 +\phi_8}{\sqrt{3}} 
            =-\phi_3+\eta_{ns} \nonumber \\
     \phi_s&=&\sqrt{\frac{2}{3}}\phi_0-\frac{2 \phi_8}{\sqrt{3}}
            =\sqrt{2}\eta_s.
\end{eqnarray}
The neutral physical states $\pi^0,\eta,\eta'$ are related to the $3\times 3$ symmetric pseudoscalar meson mass matrix of elements $B_{ij}$ emerging in the $i,j=\{0,3,8\}$ channels of $L_{mass}$ by a sequence of two transformations ${\cal S}={\cal U}{\cal V}$ that diagonalize it
\begin{eqnarray}
\label{trans}
&& \left(\phi_3,\phi_0,\phi_8\right)S^{-1} S  \left(\begin{array}{ccc} B_{33} & B_{03} & B_{38} \\
                            B_{03} & B_{00} & B_{08} \\
                            B_{38} & B_{08} & B_{88} \\  
   \end{array} \right)S^{-1} S \left(\begin{array}{c} \phi_3 \\ \phi_0 \\ \phi_8 \\
         \end{array} \right), \nonumber \\
\end{eqnarray}
first a rotation to the strange-nonstrange basis through the orthogonal involutory matrix ${\cal V}$
\begin{eqnarray}
\label{sns}
&&\left(\begin{array}{c} \phi_3 \\ \eta_{ns} \\ \eta_s
         \end{array} \right)= {\cal V} \left(\begin{array}{c} \phi_3  \\ \phi_0 \\ \phi_8
         \end{array} \right)  
\end{eqnarray}
with 
\begin{eqnarray}
\label{V}
&& {\cal V}= \frac{1}{\sqrt{3}}\left(\begin{array}{ccc} \sqrt{3} & 0 & 0 \\
                            0 & \sqrt{2}  & 1 \\
                            0 & 1 & -\sqrt{2} \\  
  \end{array} \right),
\end{eqnarray}
and then through the unitary transformation ${\cal U}$ to the physical states \cite{Kroll:2005}
\begin{equation}
\label{lin}
    \left( \begin{array}{c} \pi^0 \\
                            \eta  \\
                            \eta'
           \end{array} \right)
    = {\cal U}(\epsilon_1,\epsilon_2,\psi )
\left( \begin{array}{c} \phi_3 \\
                        \eta_{ns}  \\
                        \eta_{s}
           \end{array} \right),
\end{equation}
where
\begin{equation}
\label{Uma}
   {\cal U} = \left(\begin{array}{ccc} 1& \epsilon_1 +\epsilon_2\cos\psi
                                & -\epsilon_2\sin\psi \\
            -\epsilon_2-\epsilon_1\cos\psi& \cos\psi& -\sin\psi \\
            -\epsilon_1\sin\psi & \sin\psi& \cos\psi
            \end{array} \right)
\end{equation}
The conventional definitions $\epsilon =\epsilon_2 +\epsilon_1\cos\psi,
\epsilon'=\epsilon_1\sin\psi$ for the mixing angles are used in the tables.
The unitary matrix ${\cal U}$ has been linearized in the $\pi^0-\eta$ and $\pi^0-
\eta'$ mixing angles $\epsilon_1, \epsilon_2\sim {\cal O}(\delta ), \delta
\ll 1$. This can be done because $\phi_3$ couples weakly to the $\eta_{ns}$ and $\eta_s$
states, decoupling in the isospin limit, while the mixing for the $\eta-\eta'$ system is
strong.  Nevertheless we have tested numerically the linearization by obtaining also the exact Euler angles associated with the transformation (\ref{lin}), \cite{eQCD:2016}, the differences lying within the one to two percent level for the cases studied.  

In the isospin limit \
 ${\cal U}$ in (\ref{Uma}) leads to the $2\times 2$ orthogonal tranformation $R_\psi$
\begin{equation}
\label{basisnss}
   \left(\begin{array}{c} \eta \\ \eta'
         \end{array} \right)
   =R_\psi
   \left(\begin{array}{c} \eta_{ns} \\ \eta_s
         \end{array} \right),
%   =R_{\bar\psi}
%   \left(\begin{array}{c} -X_s   \\ X_{ns}
%         \end{array} \right),
\end{equation}
with
\begin{equation}
   R_\psi =
   \left(\begin{array}{cc} \cos\psi & -\sin\psi \\
                           \sin\psi & \cos\psi
   \end{array} \right),
\end{equation}
We remind that the angle $\psi$ is related to the mixing angle $\theta_p$
\begin{equation}
\label{singoct}
          { \phi_R^0 \choose \phi_R^8 } = \left( 
          \begin{array}{cc}
          \cos\theta_p & -\sin\theta_p \\
          \sin\theta_p & \cos\theta_p 
          \end{array}
    \right)
          { \eta' \choose \eta }\, . 
\end{equation}
of the physical states $\eta,\eta'$ in the singlet-octet renormalized basis states $\phi_R^0,\phi_R^8$ as $\psi =\theta_p +\mbox{arctan}\sqrt{2}$, with the principal value of the angle $\theta_p$ comprised in the interval $-(\pi /4)\le\theta_p \le (\pi /4)$, for details please see Appendix B of \cite{Osipov:2004b} and \cite{Osipov:2006a}.

We emphasize that one can freely choose among the different orthogonal bases to address the mixing of states. The reason  it is convenient to adopt the strange-nonstrange basis is that it allows to infer whether the mixing-parameters are determined in a process independent way;  in the context of ChPT it has been shown to be so if certain OZI violating processes are suppressed, \cite{Feldmann:1999},\cite{Kroll:2005}\cite{Kaiser:2000}, and with the exception of those originating from topolgical effects due to the $U(1)_A$ anomaly.  At this level of accuracy the decay constants follow the pattern of particle state mixing in this basis which is tantamount to having a single mixing angle involved in the determination of the decay constants associated with the $\eta$ and $\eta'$ mesons. We show in the next subsection that our model fulfills this condition at the approximation considered.
% In this case there is a single mixing angle $\vartheta_{ns}= \vartheta_s=\psi$, please see (\ref{mixsns}) below, involved in %the determination of the , which is also the outcome within our model. %Once the physical decay constants will be available %with more precision, the departure from the OZI rule can be quantified by extracting the difference in the mixing %angles. 
%In the case that isospin breaking is considered the additional mixing angles $\epsilon, \epsilon'$ are related to $\pi^0-\eta$ %and $\pi^0-\eta'$ mixing.

At this point one should remind the reader how the model's $m_{\eta'}$ contains a term related with the topological vacuum susceptibility. 
 The generalized NJL Lagrangian which combines the $U_A(1)$ breaking by the 't Hooft ($2N_f$) determinantal Lagrangian  with the 4q and 8q interactions has been shown in \cite{Osipov:2006a} to be in correspondence with the  Witten-Veneziano formula \cite{Veneziano:1979} which relates $m_{\eta'}$, in  the large $N_c$ limit of QCD with massive  quarks,  to the topological susceptibility  ${ \chi}_{YM}$ of pure Yang-Mills,  
\begin{equation}
m_{\eta'}^2 +m_{\eta}^2 -2 m_K^2= - \frac{6}{f_{\pi}^2} \chi_{YM}.
\end{equation}
In our model the topological susceptibility is obtained in the large $N_c$ limit as the following combination of model parameters
\begin{equation}
 \chi_{YM}= \frac{\kappa}{4} (\frac{M}{2G})^3.
\end{equation}
 The cutoff $\Lambda$, which is the approximate scale at which dynamical chiral symmetry breaking sets in, does not appear explicitely in the relation for $\chi_{YM}$, only hidden in the constituent quark mass M. This expression shows a judicious interplay of the subleading in $N_c$ counting $ U_A(1)$ breaking parameter $\kappa$ and the LO in $N_c$ 4q coupling G, which combine in a relation that survives in the large $N_c$ limit. This shows that the roles of chiral symmetry breaking and the breaking through the Adler Bell Jackiw anomaly are intertwined to equip $\eta'$ with its large mass, which does not vanish in the chiral limit. 
\vspace{0.5cm}

{\bf III b. Decay parameters}

\vspace{0.5cm}
At the order of the heat kernel considered, we obtain the model's axial-vector current as
 \cite{Osipov:2006a} {\footnote{The isospin current is conserved in the flavor subspace $\{0,3,8\}$ and is of no further interest for the $\pi^0-\eta-\eta'$ system discussed. }}
\begin{eqnarray}
\label{axialcur}
&&{\cal A}^a_\mu=  \nonumber \\
&&\frac{1}{4} {\mbox tr} [(\{\sigma^R+ M g^{-1},\partial_\mu \phi^R\}-\{\partial_\mu \sigma^R,\phi^R\})\lambda_a] 
+{\cal O}(b_3) \nonumber \\
\end{eqnarray}
in terms of the anti-commutators involving the bosonized and renormalized fields $\sigma^R,\phi^R$ (\ref{ren}) and the constituent quark mass matrix $M=M_a \lambda_a$. From here it is straightforward to calculate the  matrix elements in the singlet-octet basis
\begin{equation}
<0|{\cal A}^a_\mu(0)|\phi^b_R>=i f^{ab} p_\mu.
\end{equation}
One has
\begin{eqnarray}
&&f^{00}=\frac{M_u+M_d+M_s}{3g}, \hspace{0.3cm} f^{11}=f^{22}=f^{33}=\frac{M_u+M_d}{2g} \nonumber \\
&&f^{88}=\frac{M_u+M_d+4M_s}{6g}, \hspace{0.3cm} f^{08}=\frac{M_u+M_d-2M_s}{3\sqrt{2} g} \nonumber \\
&&f^{03}=\sqrt{2} f^{38}=\frac{M_u-M_d}{\sqrt{6} g} \\
%\hspace{0.3cm} f^{38}=\frac{M_u-M_d}{2\sqrt{3} g} \\
&&f^{44}=f^{55}=\frac{M_u+M_s}{2 g}, \hspace{0.1cm}  f^{66}=f^{77}=\frac{M_d+M_s}{2 g}. 
\end{eqnarray}
In particular one obtains  $<0|{\cal A}^{1+i2}_\mu(0)|\pi(p)>=i \sqrt{2} f_\pi p_\mu$ and 
$<0|{\cal A}^{4+i5}_\mu(0)|K(p)>=i \sqrt{2} f_K p_\mu$
with $f_\pi=\frac{M_u+M_d}{2 g}$ and $f_K=\frac{M_u+M_s}{2 g}$, at the order of the heat kernel expansion considered.

The neutral axial vector currents can alternatively be taken in the strange non-strange basis (\ref{dico})
\begin{eqnarray}
&&{\cal A}^{ns}_{\mu}=\sqrt\frac{2}{3}{\cal A}^0_\mu +\sqrt\frac{1}{3}{\cal A}^8_\mu, \nonumber \\ 
&&{\cal A}^{s}_\mu=\sqrt\frac{1}{3}{\cal A}^0_\mu -\sqrt\frac{2}{3}{\cal A}^8_\mu ,
\end{eqnarray}
for which one obtains the decay constants
\begin{equation}
\label{decsns}
<0|{\cal A}^{\sigma}_\mu(0)|\phi^\tau_R>=i f^{\sigma \tau} p_\mu, \hspace{0.5cm} \{\sigma,\tau\}={3,ns,s}
\end{equation}
\begin{eqnarray}
&&f^{ns}=f^{3}=\frac{M_u+M_d}{2g}, \hspace{0.3cm} f^{s}=\frac{M_s}{g}, \hspace{0.3cm} \nonumber \\ &&f^{3,ns}=\frac{M_u-M_d}{2g}, \hspace{0.3cm} f^{3,s}=f^{ns,s}=0,
\end{eqnarray}
where $f^{3},f^{ns},f^{s}$ are short-hand notations for $f^{3,3},f^{ns,ns},f^{s,s}$. The elements of (\ref{decsns})
are collected in the following matrix
\begin{equation}
\label{Fsns}
   {\cal F} = \left(\begin{array}{ccc} f_\pi & z f_\pi
                                & 0 \\ z f_\pi &  f_\pi  & 0 \\
                                0 & 0 & f_s
            \end{array} \right)
\end{equation}
with $z=\frac{M_u-M_d}{M_u+M_d}=\frac{f^{3,ns}}{f_\pi}$ marking the departure from isospin symmetry. It is of the order of the ratio involving  $\frac{f_u-f_d}{f_u+f_d} \sim {\cal O}(\delta)$, with $f_f=M_f/g$  the decay constants  ${\cal F}_f=diag\{f_u,f_d,f_s\}$.

According to the idea behind the strange-nonstrange basis, the transformation to obtain the physical decay constants 
\begin{equation}
\label{df} 
{\cal F}_P={\cal  U} {\cal F}, \quad P=\{\pi^0,\eta,\eta'\},
\end{equation}
is the same that transforms the states, eq.(\ref{lin}). In the following we discuss how the obeservables obtained from our model Lagrangian fulfill this condition.
%The first condition tells that
%\begin{equation}
%{\cal F}_P= U {\cal F}
%\end{equation}
In order to do so,  it is convenient to express the meson mass Lagrangian ${\cal L}_n$ of the neutral states in the flavor basis
\begin{eqnarray}
{\cal L}_n&=& \sum_{i=u,d,s}\left \{ \frac{1}{2}\left [(\partial_\mu \sigma_{iR})^2 +(\partial_\mu \phi_{iR})^2 +{\cal C}_i \sigma^2_{iR}  +{\cal B}_i  \phi^2_{iR}\right ] \right. \nonumber \\
&+&\left. \sum_{j=u,d,s} \left (\xi_{ij} \sigma_{iR} \sigma_{jR} + \zeta_{ij} \phi_{iR} \phi_{jR}\right )\right \}
\end{eqnarray}
with
\begin{eqnarray}
&&{\cal C}_i = \frac{N_c I_0 }{2 \pi^2}g^2 -\frac{2}{3} \xi_i, \quad {\cal B}_i= \frac{N_c I_0 }{2 \pi^2}g^2 -\frac{2}{3} \zeta_i, \nonumber \\
&& \quad \xi_{ij}=\frac{ g^2}{2} h^{(1)}_{ab}e_{ai}e_{b_j}, \quad \zeta_{ij}= \frac{g^2}{2} h^{(2)}_{ab}e_{ai}e_{b_j}, 
\end{eqnarray}
where $\xi_{ij},\zeta_{ij}$ are symmetric quantities. The quantities $\zeta_i$ are 
\begin{equation}
\label{zei}
\zeta_i=2 M_i^2 -M_j^2-M_k^2, \quad i\ne j\ne k.
\end{equation}
Inserting  them in ${\cal B}_i$ and using the gap equations (\ref{gap}), one obtains
\begin{equation}
\label{sim}
{\cal B}_i=g^2 \frac{h_i}{M_i}.
\end{equation} 
Similar expressions can be obtained for the $\xi_i,{\cal C}_i$, but they are not needed in the following.
The divergence of the axial current (\ref{axialcur}) reads in this basis
\begin{eqnarray}
\partial_\mu {\cal A}_\mu^i&=&(\sigma_{iR}+\frac{M_i}{g} ) \partial^2 \phi_{iR}  -(\partial^2 \sigma_{iR}) \phi_{iR}.
\end{eqnarray}
Using now the equations of motion for the $\sigma_{iR},\phi_{iR}$ fields
\begin{eqnarray}
0 &=&\partial^2 \sigma_{iR} - {\cal C}_i \sigma_{iR} -(\xi_{ij} \sigma_{jR}+\xi_{ij}\sigma_{iR}) \nonumber \\
0 &=&\partial^2 \phi_{iR} - {\cal B}_i \phi_{iR} -(\zeta_{ij} \phi_{jR}+\zeta_{ij}\phi_{iR} )
\end{eqnarray}
one obtains for the matrix elements  
\begin{equation} 
<0|\partial_\mu {\cal A}_\mu^i|\phi_{jR}>=\frac{ M_i}{g} ({\cal B}_i \delta_{ij} +2\zeta_{ij}),
\end{equation} 
which encode all the chiral and $U_A(1)$ symmetry breaking terms.
The off-diagonal elements of the inverse matrix pertinent to $\zeta_{ij}$ are calculated to be
%\begin{eqnarray}
%&&\zeta^{-1}_{us}=\frac{1}{4}(h_d \kappa +2 m_d \kappa_2 +2 g_8 m_s m_u)\nonumber \\
%&&\zeta^{-1}_{ds}=\frac{1}{4}(h_u \kappa +2 m_u \kappa_2 +2 g_8 m_s m_d) \nonumber \\
%&&\zeta^{-1}_{du}=\frac{1}{4}(h_s \kappa +2 m_s \kappa_2 +2 g_8 m_u m_d). \nonumber \\ 
%\nonumber
%end{eqnarray}
\begin{equation}
\label{zetuds}
\zeta^{-1}_{ij}=\frac{1}{4}(h_k \kappa +2 m_k \kappa_2 +2 g_8 m_i m_j), \quad i\ne j\ne k.
\end{equation}
These contributions violate the OZI rule,  the ones proportional to $\kappa,\kappa_2$ have their origin in the $U_A(1)$ anomaly. The term $\sim g_8$ is a contribution of the order of the square  of the current quark masses. 

For comparison the transition elements of the divergence of the axial current in the strange non-strange basis are
\begin{equation} 
Q^i_j=<0|\partial_\mu {\cal A}_\mu^i|\phi_{jR}>,\quad  i,j=\{3,ns,s\}
\end{equation} 
\begin{eqnarray}
\label{divac}
&&Q^3_3= \frac{f_\pi}{2}({ b}^{+}+z ({ b}^{-}-{2\zeta}_{ud})) \nonumber \\
&&Q^3_{ns}= \frac{f_\pi}{2}({ b}^{-}+z ({ b}^{+}+{2\zeta}_{ud})) \nonumber \\
&&Q^3_s= \frac{f_\pi}{\sqrt{2}} ({\zeta}_{us}-{\zeta}_{ds}+z({\zeta}_{us}+{\zeta}_{ds}))
 \nonumber \\
&&Q^{ns}_{3}= \frac{f_\pi}{2}({ b}^{-}+z ({ b}^{+}-{2\zeta}_{ud})) \nonumber \\
&&Q^{ns}_{ns}= \frac{f_\pi}{2}({ b}^{+}+z ({ b}^{-}+{2\zeta}_{ud})) \nonumber \\
&&Q^{ns}_s= \frac{f_\pi}{\sqrt{2}} ({\zeta}_{us}+{\zeta}_{ds}+z({\zeta}_{us}-{\zeta}_{ds}))\nonumber \\
&&Q^{s}_3= \frac{f_s}{\sqrt 2}({\zeta}_{us}-{\zeta}_{ds}),  \quad  Q^{s}_{ns} = \frac{f_s}{\sqrt 2}({\zeta}_{us}+{\zeta}_{ds}) \nonumber \\
&&Q^s_s= f_s (\frac{{\cal B}_s}{2} +\zeta_{ss}) \
\end{eqnarray}
with
\begin{eqnarray}
&&b^{\pm}=(\frac{{\cal B}_u}{2}+\zeta_{uu})  {\pm}  (\frac{{\cal B}_d}{2} +\zeta_{dd}) \nonumber \\
%&&{\bar \zeta}_{ij}=\zeta_{ij}+\zeta_{ji}, \quad {i\ne j}, \quad {i,j}=\{u,d,s\} 
\end{eqnarray}
Note that  the elements $Q^i_j$,  $i,j=\{3,ns,s\}$ are not symmetric under exchange of $\{i,j\}$ differing by terms $\sim z$ and $\sim y=\frac{f_s}{f_\pi}$.
Obviously in the isospin limit the elements $Q^3_{ns},Q^{ns}_3,Q^3_{s},Q^s_3$  vanish. In this limit the $Q^s_{ns} =y Q^{ns}_s $, $y$ being a measure for flavor breaking in the light vs. strange sector.

Finally using the relations (\ref{lin}),(\ref{Uma}) and (\ref{divac})  we calculate the vacuum to physical particle transitions of the divergence of the axial current,  $<0|\partial_\mu {\cal A}_\mu^b|P>, \quad  b=\{3,ns,s\}, \quad P=\{\pi^0,\eta, \eta'\}$,  discarding terms  $\sim \delta^2$,  and are able to  show that it fulfills the following relation \cite{Kroll:2005}
\begin{equation}
\label{decpat}
 <0|\partial_\mu {\cal A}_\mu^b|P_a>= {\cal M}^2_{aa'} {\cal U}_{a'b'} {\cal F}_{b'b}.
\end{equation}
with $ {\cal F}$ given by (\ref{Fsns}) and ${\cal M}_{aa'}$ the physical meson mass matrix, which on the requirement of being diagonal determines the mixing angles $\psi,\epsilon,\epsilon'$. Thus the decay constants transform as the states (\ref{lin}) within our model calculations, at the order of the heat kernel considered. Higher order terms involve derivative interactions, which are likely to change this behavior. 

We obtain the follwing relations at ${\cal O}(\delta)$
\begin{eqnarray}
\label{decsqua}
&&\sum_{P=\pi^0,\eta,\eta'}f_P^3f_P^3=f_\pi^2, \quad \sum_{P=\pi^0,\eta,\eta'}f_P^{ns}f_P^{ns}=f_\pi^2 \nonumber \\
&&\sum_{P=\pi^0,\eta,\eta'}f_P^s f_P^s=f_s^2, \nonumber \\
\end{eqnarray}
for the diagonal elements, and
\begin{eqnarray}
\label{decnd}
&&\sum_{P=\pi^0,\eta,\eta'}f_P^{ns} f_P^s=\sum_{P=\pi^0,\eta,\eta'}f_P^3 f_P^s=0, \nonumber \\
&&\sum_{P=\pi^0,\eta,\eta'}f_P^3 f_P^{ns}=z f_\pi^2
\end{eqnarray}
for the crossed terms. The last relation is a consequence of (\ref{Fsns}). The vanishing of the other crossed terms indicates that just one mixing angle is present for the $\eta-\eta'$ mixing in the strange nonstrange basis at this order of approximation.  This can be  seen in a simple way in the isospin limit, and the argument is the same in the general case.
In the isospin limit the physical states $P=\eta, \eta'$ are now only mixtures of the nonstrange
and strange components, as follows from  (\ref{basisnss}).
This case has been considered in \cite{Osipov:2006a} (where a detailed discussion of the mixing scheme in connection with the singlet octet basis in the isospin limit is also presented),
\begin{equation}
   \langle 0|{\cal A}_\mu^{i}(0)|P(p)\rangle =if_P^i p_\mu, \qquad
   (i=ns,s).
\end{equation} 
The couplings can be represented in a way which is similar to the one of 
Leutwyler -- Kaiser \cite{Feldmann:1999},\cite{Kaiser:2000}, who introduce two mixing angles $\vartheta_{ns},\vartheta_{s}$ 
\begin{equation}
   \{f^i_P\} =
   \left( \begin{array}{cc}
          f^{ns}_\eta & f^s_\eta \\
          f^{ns}_{\eta'} & f^s_{\eta'}
          \end{array}
   \right) = \left( 
          \begin{array}{cc}
          f_{ns}\cos\vartheta_{ns} & -f_s\sin\vartheta_s \\
          f_{ns}\sin\vartheta_{ns} & f_s\cos\vartheta_s 
          \end{array}
    \right)
\end{equation}
Our calculations show that within our model,   with ${\bar M}=M_u=M_d$, 
\begin{eqnarray}  
   && f^{ns}_{\eta}=\frac{{\bar M}}{g}\cos\psi , \qquad
      f^{s}_{\eta}=-\frac{M_s}{g}\sin\psi , \nonumber \\
   && f^{ns}_{\eta'}=\frac{{\bar M}}{g}\sin\psi , \qquad
      f^{s}_{\eta'}=\frac{M_s}{g}\cos\psi . 
\end{eqnarray}
One thus obtains the relation for the mixing angles 
\begin{equation}
\label{mixsns}
\sum_{P=\eta,\eta'} f^{ns}_P f^{s}_P=f^{ns} f^{s} sin(\vartheta_{ns}-\vartheta_{s})=0.
\end{equation}
This follows as the basic parameters $f_{ns}, f_s, \vartheta_{ns}, 
\vartheta_s$ of the matrix $\{f^i_P\}$, expressed in terms of
model parameters (in the approximation considered)
\begin{equation}
   f_{ns}= \frac{\bar M}{g}= f_\pi , \quad f_s=\frac{M_s}{g}\, , \quad 
   \psi =\vartheta_{ns}=\vartheta_s.
\end{equation}
There is a direct relation between the common mixing angle
$\vartheta_{ns}=\vartheta_s$ and the OZI-rule which has been discussed 
in \cite{Feldmann:1999},\cite{Kaiser:2000}. 
The model predictions agree well with the general
requirements of chiral symmetry following from chiral perturbation 
theory (ChPT), although the results differ already at lowest order. For 
instance, we have
\begin{equation}
\label{fsiso} 
f_s^2=2{\bar f}_K^2 -{\bar f}_\pi^2 +\frac{(M_s-{\bar M})^2}{2 {\bar g}^2}
\end{equation}
where barred quantities are a reminder that they are taken in the isospin limit.
 The first two terms
of this formula are a well known low-energy relation which is valid in 
standard ChPT. 
In the $\eta'$-extended version of ChPT there is a OZI-rule violating
term in the effective Lagrangian, which contributes as 
${\bar f}_\pi^2\Lambda_1$ to r.h.s. of (\ref{fsiso}) . We have   
instead the term $(M_s-{\bar M})^2/(2{\bar  g}^2)$. Of course, in our
case the origin of this contribution is related with the $SU(3)$
flavour symmetry breaking effect and does not have impact on the deviation from a single mixing angle in the strange-nonstrange basis reported as consequence of the OZI-rule violating parameter $\Lambda_1$.

Away from the isospin limit one obtains  
\begin{equation}
\label{fsg}
 f_s^2=2 f_K^2  - f_\pi^2(1+2z) +\frac{(M_s-M_u)^2}{2 g^2},
\end{equation}
with a correction of order $\delta$ in the  $f_\pi^2$ term, as compared to (\ref{fsiso}). The numerical values that we obtain are $z \sim 4\times 10^{-3}$, which amounts to a small correction of $\sim 0.8 \%$ in the $f_\pi^2$ term. 

\vspace{0.5cm}

For completeness we indicate the decay constants and mixing angles in the singlet-octet basis in the isospin limit, used in Table (\ref{0-8angles}), for details see please \cite{Osipov:2006a}, 
\begin{eqnarray}
\label{0-8ang}
\theta_0&=&\psi-arctan(\sqrt 2 \frac{\bar M}{M_s}), \nonumber \\
\theta_8& =&\psi-arctan(\sqrt 2 \frac{M_s}{\bar M}) \nonumber \\
\psi &=&\theta_P  +arctan \sqrt 2
\end{eqnarray}
\begin{eqnarray}
\label{f0f8}
f_0^2&=&\frac{2{\bar f_K}^2+{\bar f_\pi}^2}{3} +\frac{{\bar f_\pi}^2}{6}(\frac{M_s}{\bar M}-1)^2 \nonumber \\
f_8^2&=&\frac{4{\bar f_K}^2-{\bar f_\pi}^2}{3} +\frac{(M_s-{\bar M})^2}{3{\bar g}^2}
\end{eqnarray}

{\bf IV. Results and further discussion} 
\vspace{0.5cm}

The model has 15 parameters, 4 couplings $G,\kappa,g_1,g_2$ associated with $L_{int}$ in (\ref{L-int}), 7 non-vanishing couplings  $\kappa_2, g_3....g_8$ in in the ESB sector (\ref{L-chi-1}), the cutoff $\Lambda$, and the 3 current quark masses. Before running a fit it is convenient to understand to which parameters the difference of the light quark masses $(m_u-m_d)$ is most sensitive. For that it is instructive to look at the % $(h_{03}^{(2)})^{(-1)},(h_{38}^{(2)})^{(-1)}$%  
matrix components (\ref{h2}) in the strange-nonstrange basis, which vanish in the isospin limit.  As mentioned before these elements belong to the SPA contribution to the meson mass Lagrangian, see last line in (\ref{mass}), which is the part that carries the full information on the model couplings. Note that the heat kernel contribution to the meson masses, represented except for the kinetic terms in the remaining of expression (\ref{mass}), only depends on the cutoff $\Lambda$ of the $I_i$ quark integrals. The dependence on the model couplings of the heat kernel contribution only enters implicitly through the constituent quark masses, via the gap equations (\ref{gap}) which are solved selfconsistently with the lowest order SPA equations (\ref{h}). 
Defining %$M_\Delta=\frac{1}{2}(M_d-M_u)$, $M_\Sigma=\frac{1}{2}(M_d+M_u)$, 
$m_\Delta=\frac{1}{2}(m_d-m_u)$, $m_\Sigma=\frac{1}{2}(m_d+m_u)$, $h_\Delta=\frac{1}{2}(h_d-h_u)$ and $h_\Sigma=\frac{1}{2}(h_d+h_u)$, one has for the inverse matrix elements of the $\zeta_{ij}$  matrix, $i,j=\{3,ns,s\}$
%1/4 (hdel (2 g2sa hsig + g3sa sig) +  del (g3sa hsig - 2 (g5sa - g6sa + 2 g8sa) sig))
\begin{eqnarray}
&&(\zeta_{3,ns})^{(-1)}=\frac{1}{4}[h_\Delta (2 g_2 h_\Sigma + g_3 m_\Sigma)  \nonumber \\
&& + m_\Delta (g_3 h_\Sigma - 2 (g_5 - g_6 + 2 g_8) m_\Sigma)]  \nonumber \\
&& \nonumber \\
&&(\zeta_{3,s})^{(-1)}=\frac{1}{2 \sqrt{2}}(h_\Delta \kappa + 2 m_\Delta (\kappa_2 - g_8 m_s)) \nonumber \\
&& \nonumber \\
&&(\zeta_{ns,s})^{(-1)}=\frac{1}{2 \sqrt{2}}(h_\Sigma \kappa + 2 (\kappa_2 + g_8 m_s) m_\Sigma).
\end{eqnarray}
The first two elements vanish in the isospin symmetric case. The third element relates to the mixing in $\eta-\eta'$ and is non-zero in this limit. 
In this basis one sees that the mixing in the $3,ns$ sector involves other couplings as in the strangeness related sectors,  in contrast to the off-diagonal matrix elements in the $u,d,s$ basis, (\ref{zetuds}). This sets new constraints  on these couplings, as compared to the isospin symmetric case. It follows that the interplay of these parameters (which is conditioned by the fits), not the actual magnitude of each of the terms, is relevant to obtain the size of isospin corrections.  We remark  that in $(\zeta_{ns,s})^{(-1)}$ isospin breaking effects are absent, and that in the diagonal elements (not shown) they are  overshadowed by the presence of $m_s,m_\Sigma,h_s,h_\Sigma$.

The current quark mass dependence in these expressions enters together with the ESB couplings. In their absence the effects of ESB come only through the difference in the light condensates $h_\Delta$ which do not vanish if the conventional QCD mass term has $m_u\ne m_d$ values, and if the couplings $\kappa$ and $g_2$ are not zero (if they also vanish, only the heat kernel contribution to the meson mass matrix carries the effects of isospin breaking). The coupling $\kappa$ is strongly correlated with the $\eta-\eta'$ mass splitting and it enters in the corresponding $(\zeta_{ns,s})^{(-1)}$ matrix element as factor of $ h_\Sigma$ which remains approximately constant. Thus this parameter is not expected to vary much in the fit of isospin breaking effects, which has been also verified numerically. 
\vspace{0.5cm}

Before showing the results for isospin breaking, we display relevant model observables in the isospin limit, in comparison to other approaches. We consider the cases in which the ESB terms are present in the interaction Lagrangian, sets (a,b) in the Tables \ref{table-1}-\ref{0-8angles}, %\ref{explicitSB}
 and compare with the parameter set (c) in which explicit symmetry breaking occurs only through the LO current quark mass term.  Table I  indicates the mass spectra of the low lying pseudoscalar (and scalar meson nonets, see caption, for sets (a,b)) used in the fit of parameters. Sets (a,b) have different mixing angles for the pseudoscalars as well as for the scalars, leaving however the  mass spectra and weak decay constants $f_\pi,f_K$ inchanged. This is possible because the fit is done simultaneously in both sectors, leading to a readjustment of parameters.  A comment on the scalar mass spectrum:  we  were able to obtain a reasonable simultaneous fit for the $\sigma$ mass and its large decay width, at tree level of the mesonic Lagrangian. It is our understanding that this results partly from the fact that already at mesonic tree level there are signatures of quark-antiquark as well as of admixtures of 2 quark- 2 antiquark states present, from the underlying multiquark Lagrangian, which contribute in the long distance asymptotics to the internal structure of the mesons as well as to the mesonic interaction terms.  In many other approaches the more complex quark structures enter only through the explicit inclusion of  meson loops, tetraquark configurations, and so on  \cite{Jaffe:1977,Black:1999,Wong:1980,Narrison:1986,Beveren:1986, Latorre:1985,Alford:1998,Achasov:1984,Isgur:1990,Schechter:2008,Schechter:2009,Close:2002,Klempt:2007}.

Table II shows the current and constituent quark masses in the isospin limit and the model parameters which do not break explicitly the chiral symmetry, the 4-quark coupling $G$, the 6-quark �t Hooft determinant coupling $\kappa$, and  two 8-quark couplings $g_1,g_2$, of which $g_1$ is OZI-violating. In set (c) we put $g_1=2500$ without loss of generality, since a change in $g_1$ can be counterbalanced by a change in $G$ leaving all other parameters and observables unchanged, except for the low lying $\sigma$-meson mass, see e.g. \cite{Osipov:2006a}, \cite{Hiller:2010}, which does not affect the pseudoscalar characteristics considered here.  Table III displays the couplings related with the ESB interactions. 

In Table \ref{0-8angles}  further properties of the $\eta-\eta'$ system are presented, the angle $\theta_P$ in the singlet-octet basis and the related angles $\theta_0,\theta_8$, as well as the weak decays $f_0,f_8$, see eqs. (\ref{0-8ang}),(\ref{f0f8}) obtained following the methods in sections 3.2 and 3.3 of \cite{Osipov:2006a},  and compared with the results of other approaches \cite{Feldmann:1998} and references therein, \cite{Goity:2002}, \cite{Kaiser:2000},\cite{de Fazio:2000}, and of Pad\'e approximant method to pseudoscalar transition form factors \cite{Escribano:2014}, \cite{Escribano:2015},  where in the Table \ref{0-8angles}, $(\eta)$ and $(\eta')$ refer to the asymptotes taken for the fits, and the results obtained from BABAR data \cite{Babar} by the same authors (see discussion in  \cite{Escribano:2015}  concerning the BABAR "puzzle").

This comparison shows that set (a) yields results quite close to the chiral perturbation analysis of \cite{Goity:2002} and follows the general trend of the other references in the table, with exception of the sum rules approach of \cite{de Fazio:2000}, and  the fits from  \cite{Babar}. Our values for $\theta_0,\theta_8$ are smaller, respectively larger than the ones presented in  \cite{Feldmann:1998}, which is probably due to the different way in which the $U(1)_A$ anomaly is treated. One further notices that an increase in $\theta_P$ leads mainly to an increase in the $\theta_0$ angle, comparing sets (a) and (b), which have similar theoretical input. 
\vspace{0.5cm}

Turning now to isospin breaking, in the numerical fit we keep the cutoff $\Lambda$, which sets approximately the scale of chiral symmetry breaking, close to the isospin limit value, and take for the strange current quark mass the value $m_s=95$MeV. 
We do not consider the scalar spectrum in this case, the related isospin breaking effects will be addressed in a future work. Then the general case, with the ESB terms, involves  solving self-consistently a system of 13 equations, the three gap equations (\ref{gap}) subject to the three SPA conditions (\ref{h}), 6 equations for the meson mass matrix elements, that is 3 for ${m_{\pi^0}}$, ${m_{\eta}}$, ${m_{\eta'}}$ and the remaining 3 for diagonalization, one equation for the kaon mass, 2 equations for the weak decay constants $f_\pi,f_K$, one equation which fixes the ratio $r=\frac{m_d-m_u}{m_u+m_d}$ of current quark masses. We vary externally the values of the mixing angles $\epsilon,\epsilon',\psi$ and search for the parameters that lead to the best fit of the $m_u,m_d$ current masses. The result is indicated in Table VII, using  empirical input and couplings shown in Tables V and VI (the empirical splitting of the charged multiplets cannot be reproduced, since electromagnetic effects are not taken into account). In presence of ESB (sets A and B) with the ratio $r$ kept equal and close to its empirical value, the ratio $\frac{\epsilon}{\epsilon'}$ is well reproduced in comparison with the literature shown in Table VIII. The main observation is that this ratio is reduced by $\sim 40 \%$ compared to the model variant without ESB interactions, set C. In the latter case one does not obtain $\frac{\epsilon}{\epsilon'}$ nor $m_u,m_d$  close to the empirical values. Neither this ratio nor the light current quark masses get improved in set C by reducing $r$ down to $0.2$, the main consequence being a drastic change in the values for the mixing angles, which get reduced to $\epsilon=0.0119, \epsilon'=0.00135$. We note that it is not possible to get a better fit for $m_\eta$, $m_K$ and $f_K$ in absence of ESB interactions as the one shown in set C in Table V.
Contrary to this, the individual values for $\epsilon$ and $\epsilon'$ for sets A,B are in good agreement with the ones indicated in \cite{Feldmann:1999}, \cite{Kroll:2005} and the corresponding current quark masses are very close to the quoted values  $m_u=2.15(15)$ MeV, $m_d=4.70(20)$ MeV,  \cite{PDT:2014}.% On the other hand our ratio $m_u/m_d \sim 0.46$ is also close to Weinberg's prediction $m_u/m_d=0.56$ \cite{Weinberg:1977} that corresponds to the leading order ChPT result, showing that the NLO in $N_c$ corrections that we take into consideration represent a moderate correction to this value.   

%We first show the results for the case without the ESB interaction terms, set. For a best possible fit of the %pseudoscalar meson mass spectrum, as well as of the weak decay constants $f_\pi,f_K$, one does not obtain the empirical% ratio $\frac{\epsilon}{\epsilon'}$ nor values close to the actual current quark masses, as shown in table. 
  
Regarding table VIII that collects values obtained in the literature, within different phenomenological approaches, as well as in  experiments, a comparison of the different values has to be done with care, for a careful and detailed discussion see \cite{Kroll:2005}. We note in particular that in the experimental value \cite{Tippens:2001} the mixing $\pi^0-\eta'$ has not been taken into account and that in the ChPT result \cite{Goity:2002} the $\eta'$ is considered as a background field.

%Having shown that the parameters obtained in the isospin limit lead to a good description of empirical data, we proceed to show the results of isospin breaking in Table V. We use the same three values of $r$ in each parameter set (a,b,c). The amount by which the current quark masses differ, related to $\delta$, leads to a slight reduction in the $\pi^0$ mass, which turns out to be at most 1 MeV. The larger observed empirical value for the difference $m_{\pi^{\pm}}-m_{\pi^{0}}\sim 4.5$ MeV is mainly of electromagnetic origin, not considered here. The main observation is that the ratio $\frac{\epsilon}{\epsilon'}$ remains almost constant within each set and that the  sets (a,b) with explicit chiral symmetry breaking terms yield a ratio which is reduced by $\sim 40 \%$ compared to set (c). The larger value of $r$ leads in sets (a,b) to current quark masses of the light quarks which are closer to the presently quoted values  $m_u=2.15(15)$ MeV, $m_d=4.70(20)$ MeV,  \cite{PDT:2011}. The case denoted as set (a') corresponds to solving the gap equations with the parameter set (a) keeping the NLO current quark mass insertions in ($\ref{h}$) fixed to the value $m_0$. As mentioned above, a rescaling of $\delta$ leads to the same output for constituent masses and mixing angles. However the large value $r$ needed for that requires a much too large splitting in the $m_u,m_d$ values, as compared to empirical values.

\vspace{0.5cm}

{\bf  V. Conluding remarks}
\vspace{0.5cm}

We conclude that the explicit symmetry breaking interactions of the generalized NJL Lagrangian considered are crucial to obtain the phenomenological quoted value for the ratio $\frac{\epsilon}{\epsilon'}$. We obtain values for the $\epsilon$ mixing angle which lie within the results discussed in the literature. Unfortunately the value for $\epsilon'$ is much less discussed. We obtain $\epsilon$ and $\epsilon'$ reasonably close to the ones indicated in \cite{Feldmann:1999}, \cite{Kroll:2005} for current quark mass values in excellent agreement with the presently quoted average values. The corresponding sets (A,B) are the ones which also yield the best fits to other empirical data within the model variants. 
\vspace{0.5cm}

{\bf Acknowledgements}

We thank W. Broniowski for reading the manuscript and for critical remarks.
Work supported in part by Funda\c{c}\~ao para a Ci\^encia e
Tecnologia and Centro de F\'{\i}sica Computacional da Universidade de Coimbra.

%%%%%%%%%%%%%%%%%%%%%%%%%%%%%%%%%%%%%%%%%%%%%%%%%%%%%%%%%%%%%%%%%%%%%%%%%%%%%

\end{document}